\documentclass[]{aa}
\usepackage{graphicx}
\begin{document}     
\title{Dust temperature and density profiles in the envelopes of AGB and post-AGB
   carbon stars from mid-infrared observations\thanks{Based on observations made
   at La Silla ESO Observatory (Chile).}}

 \author{E. Lagadec\inst{1}                                      
    \and D. M\'ekarnia\inst{1}
    \and J.A. de Freitas Pacheco\inst{2}
    \and C. Dougados \inst{3}}

 \offprints{E. Lagadec, e-mail: lagadec@obs-nice.fr}
 \institute{Dpt Cassiop\'ee, CNRS-UMR 6202, Observatoire de la C\^ote d'Azur,
             BP 4229, 06304 Nice Cedex 4, France
           \and  
             Dpt Art\'emis, CNRS-UMR 6162, Observatoire de la C\^ote d'Azur, BP 4229, 
             06304 Nice Cedex 4, France 
           \and 
             Laboratoire d'Astrophysique, Observatoire
             de Grenoble, 414 Rue de la Piscine, 38400 Saint-Martin
             d'Hy\`eres, France 
            }
 \date{Received / Accepted 18 November 2004}
 \titlerunning{Dust temperature and density profiles of AGB and post-AGB stars}
 \authorrunning{Lagadec et al.}  

  \begin{abstract} 
   { 
    First mid-infrared images of a sample of AGB and
    post-AGB carbon stars (V\,Hya, IRC\,+10216, CIT\,6 and Roberts\,22)
    obtained at La Silla Observatory (ESO, Chile) are
    reported.  CIT\,6 presents a cometary-like feature clearly seen in
    the 9.7\,$\mu$m image, Roberts\,22 shows an envelope slightly elongated in the
    north-east direction  while images of V\,Hya and IRC\,+10216 are
    roughly spherically symmetric. Using inversion technique, the
    dust emissivity was derived from the observed intensity profiles,
    allowing a determination of
    the grain temperature and density distributions inside the
    envelope for these stars.  Dust masses and mass-loss rates  were 
    estimated for V\,Hya and IRC\,+10216. Our results are comparable
    to those obtained in previous studies if dust grains  have dimensions in the
    range $\sim$\,0.01\,-\,0.2\,$\mu$m. Color maps
    suggest the presence of temperature inhomogeneities in the central
    regions of the dust envelopes. In the case of V\,Hya, an eccentric 
    hot point, which direction coincides
    with the jet previously seen in [S\,II] emission, suggest that we are
    observing a material ejected in a previous mass-loss event. Bipolar
    lobes are clearly seen in the color maps of  Roberts\,22 and IRC\,+10216.
  
  } 
  \end{abstract}

  \maketitle

\section{Introduction}

According to our current understanding of stellar evolution, all stars with 
main sequence progenitors in the mass range 1\,-\,8\,${\mathrm M}_{\odot}$, evolve via the
asymptotic giant branch (AGB) phase to the planetary nebula (PN)
stage. The early AGB (E-AGB) phase begins as soon as the star finishes
burning  He in its core and begins to burn He in a thick shell.  When both  H and He burning
shells are active, the star enters in the thermally
pulsating AGB (TP-AGB) phase (Iben \& Renzini 1983; Bl\"ocker 1999).

The AGB phase is characterized by a strong mass-loss period that only stops when the
outer envelope is completely lost (Habing 1996). As the star
ascends the AGB, the mass-loss rate
increases from solar-like values ($10^{-14}\, {\mathrm M}_{\odot}/\mathrm{yr}$)  up 
to $10^{-4}\,{\mathrm M}_{\odot}/\mathrm{yr}$, the so-called
super-wind phase. Dust grains and molecules, predominantly CO, are
formed in the wind, producing a circumstellar envelope detectable in the
infrared and millimetric domains. Although the radiation pressure on dust
grains and on molecules is thought to play an important role in the
driving mechanism of the wind, different physical aspects
governing  the dust formation/destruction processes are still unclear. 
Processed elements are dredged-up into the atmosphere as a consequence of the
re-ignition of the He burning shell (E-AGB phase). A further contraction leads
also to the re-ignition of the H burning shell and the star enters a
TP-AGB phase. A third dredge-up may take place during this phase and
overshooting will contribute to lift carbon to the surface.  These events 
lead to the formation of  circumstellar envelopes
whose chemical composition reflects that of an atmosphere enriched
with nuclear processed matter. C-rich stars  have envelopes 
composed mainly of amorphous carbon
while O-rich stars have circumstellar envelopes  rich in
silicates. Both oxidic and carbonaceous dust particles are characterized by
vibrational bands positioned in the mid-IR window (at 9.7\,$\mu$m and 
18\,$\mu$m for silicates  and at 11.3\,$\mu$m for SiC) (Henning 1999). Most of
the thermal radiation continuum from the dusty  envelope is
emitted in this  wavelength range.

Mid-infrared imaging allow direct probe of the mass distribution
through thermal emission arising from dust grains. Therefore a number
of imaging studies of AGB and post-AGB stars have been made at mid-infrared
wavelengths and some emission regions have been resolved (Dayal et al. 1998; Meixner et
al. 1999; Ueta et al. 2001a,b; Jura et al. 2002a,b; Kwok et al. 2002;
Close et al. 2003; Gledhill \& Yates 2003).

In the present work we report mid-infrared imaging, with a spatial
resolution of about 0.7\,\arcsec, of a sample of AGB and post-AGB
carbon-rich stars (CIT\,6, V\,Hya, Roberts\,22 and IRC\,+10216) not previously imaged
at these wavelengths. Our goals  are to derive  physical properties of the dust
inside the circumstellar envelope and to estimate mass-loss rates.
Most of the objects  have circumstellar envelopes nearly  symmetric, allowing
the use of inversion techniques to obtain the dust emissivity and, consequently,
the density and the temperature as a function of the radial distance
from the star. This paper is organized as follows: in Section~2,
observations and data reduction are described; in Section~3,
morphological analyzes of the individual objects are presented; in
Section~4, we describe our model for the envelope and present the
derived dust density and temperature profiles; in Section~5 we discuss our results
and compare with previous works;  finally, in Section~6, the
main  conclusions are  given.

\section{Observations and data reduction}
Observations were performed on 1995 February 5, 7 at the ESO
3.6-m telescope in La Silla (Chile), using the  mid-infrared TIMMI 
camera (Lagage et al. 1993) with a detector consisting of  a $64\,\times\,64$
pixel Ga:Si array.
The resulting spatial scale of the system is 0.33\,\arcsec\,/pix.
The log of the observations is given in Table \ref{tab_log}. These observations
were made in a standard mid-IR observing mode, by
chopping the secondary mirror  and nodding the telescope
to subtract the background emission from the sky and telescope. The
chopper throw was 18.3\,\arcsec \, toward the south and the nod beam position
used was 18.3\,\arcsec \, north of the first position. To avoid the
saturation of the detector by the ambient photon background and 
to have a good image quality, each individual nod cycle was split into
many short exposures of  $\sim$\,10\,ms. This procedure was repeated
for as many cycles as needed to obtain the required total integration
time. Nearly diffraction-limited images ($\sim\,0.7\,\arcsec \,$ FWHM for point
sources) resulted from these short exposures. Filter wavelengths
were selected in order to obtain information on the dust continuum,
while maximizing the detection sensitivity. The observations were
carried out  in four narrow-band filters, centered respectively at
8.39\,(N1), 9.7\,(Silicates), 9.78\,(N2), and 11.65\,$\mu$m\,(SiC)
whose characteristics are given in Table \ref{table_filtre}. During 
the different runs the seeing was typically  0.7\,\arcsec\,-\,0.8\,\arcsec.

Data reduction was performed using IDL self-developed routines.  
Individual chopped  frames were  spatially oversampled
by a factor of 4 and shifted to the nearest 0.25 pixel by using a cross-correlation 
algorithm to correct for turbulent
motions and flexure drifts.
Images were then co-added to produce a single 
flat-field-corrected image, comprising the average of the chop and nod 
differences, for each filter. Reference stars were 
observed, analyzed in the same way and were used to derive the instrumental point spread
function (PSF) at each filter. Images were deconvolved by 
using the Maximum Entropy Method (Hollis et al. 1992) and  
fluxes were obtained by interpolation of IRAS-LRS data. Table \ref{table_flux} 
gives adopted fluxes for each observed object.

   \begin{table}
      \caption[]{Log of the observations.}
         \label{tab_log}
         \begin{tabular}[]{l l l l }
            \hline
            \hline
            \noalign{\smallskip}
            Object & Date (UT) &  Filter  \\
           \noalign{\smallskip}
            \hline
            \noalign{\smallskip}
  CIT\,6         & 1995 Feb 7 & N1, Silicates    \\  
  V\,Hya         & 1995 Feb 7 & N2, SiC   \\
  Roberts\,22    & 1995 Feb 5 & N1, SiC  \\
  IRC\,+10216    & 1995 Feb 5 & N1, SiC  \\  
            \noalign{\smallskip}
            \hline
         \end{tabular}
   \end{table}

   \begin{table}
      \caption[]{Description of the filters used.}
         \label{table_filtre}
         \begin{tabular}[]{l c c}
            \hline
            \hline
            \noalign{\smallskip}
            Filter & Central wavelength & Filter width \\
                   & ($\mu m$)          & ($\mu m$)   \\
           \noalign{\smallskip}
            \hline
            \noalign{\smallskip}
  N1         &  8.39   & 7.91\,-\,8.87     \\
  Silicates  &  9.70   & 9.50\,-\,10.00    \\
  N2         &  9.78   & 9.14\,-\,10.43   \\          
  SiC        & 11.65   & 10.3\,-\,13.00  \\
          \noalign{\smallskip}
            \hline
         \end{tabular}
   \end{table}

   \begin{table}
      \caption[]{Adopted mid-infrared fluxes for observed objects.}
         \label{table_flux}
         \begin{tabular}[]{l l c}
            \hline
            \hline
            \noalign{\smallskip}
            Object & Filter & log flux  (Jy)
              \\
           \noalign{\smallskip}
            \hline
            \noalign{\smallskip}
  CIT\,6       & N1   & 3.58  \\
               & Silicates   & 3.55  \\
  V\,Hya       & N2   & 2.85  \\
               & SiC  & 2.87 \\
  Roberts\,22  & N1   & 1.89  \\
               & SiC  & 2.28  \\
  IRC\,+10216  & N1   & 4.42 \\
               & SiC  & 4.63 \\
          \noalign{\smallskip}
            \hline
         \end{tabular}
   \end{table}

\section {Image analysis of individual objects}

   \begin{figure*}
 \centering
 \includegraphics[width=18.cm]{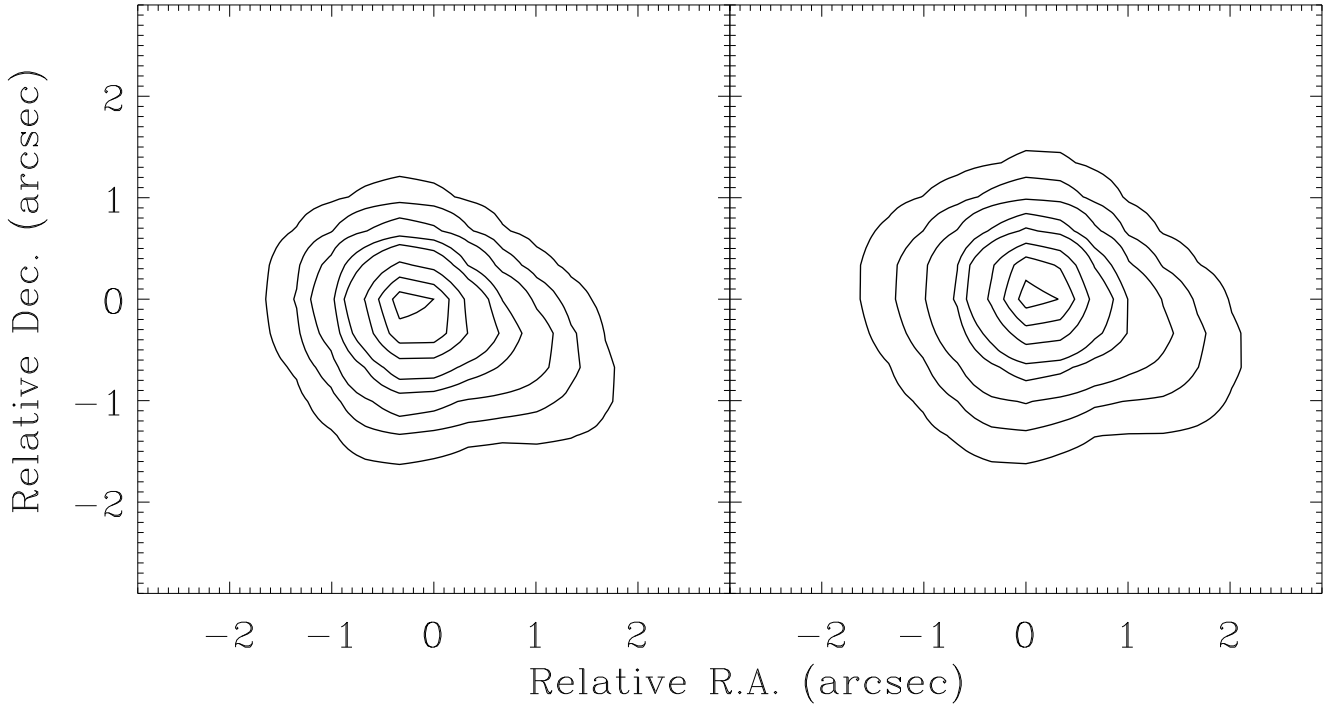}
 \caption{Contour plots of the 9.78\,$\mu\mathrm{m}$ (left) and 
   11.65\,$\mu\mathrm{m}$ (right) images of V\,Hya. North
   is up and east is to the left. Contour levels are 90, 70, 50, 30 , 20,
   10, 5 and 2\,\% of the peak surface brightness.}
  \label{FigVHya}
  \end{figure*}
 
\subsection{V\,Hya}
V\,Hya (IRAS\,10491-2059) is a N type AGB carbon star, with an
effective temperature of 2\,650\,K and a C6\,-\,7.5e spectral type (Knapp et al.
1997). It is a semi-regular variable, with a period of about 529 days
(Lambert et al. 1986), surrounded by  a
bipolar envelope as shown by  CO observations (Tsuji
et al. 1988; Kahane et al. 1996) and near-infrared polarization
data (Johnson \& Jones 1991; Trammell et al. 1994). High
resolution optical spectroscopy  suggests that V\,Hya is 
in rapid rotation, which could be induced by 
the presence of a companion (Barnbaum et
al. 1995). The large  infrared excess (van der Veen 
\& Habing 1988) as well as the  strong emission in
molecular lines  indicate that V\,Hya is losing mass at a rate higher
than $10^{-6}\,{\mathrm M}_{\sun}\,\mathrm{yr}^{-1}$. 
Knapp et al. (1997) detected, via
CO observations, a fast molecular wind ($V_w\,\sim\,200\,\mathrm{km\,s}^{-1}$) expanding from
the poles of the envelope and having an increasing velocity  with the distance
from the star. From recent HST observations, Sahai et al. (2003)  report the discovery
of a newly ejected high-speed jet-like outflow in this
star. These observations, combined with  a previous
interferometric CO (J=1-0) map of V\,Hya, favor the picture of 
an expanding, tilted and dense disk-like structure, oriented
north-south present inside the inner envelope.

We obtained images of V\,Hya at
9.78 and 11.65\,$\mu\mathrm{m}$ (N2 and SiC filters
respectively). Images are roughly spherically symmetric in the central
parts of the envelope but isophotes
become slightly elongated in the south-west direction at intensity levels
lower than 20\,\% of the central value. The 
N2-image has an extension at 2\,\% intensity level 
of $4\,\arcsec\,\times\,4\,\arcsec$ whereas in the 
SiC filter the dimension is 
$4.5\,\arcsec\,\times\,4.5\,\arcsec$ (Fig. \ref{FigVHya}).
The observed elongated feature in the south-west direction could be associated to
the dust emission from material blown away perpendicular to the equatorial
disk, consistent with the model proposed by Sahai et al. (2003).

\begin{figure}
  \centering
  \includegraphics[width=9.cm]{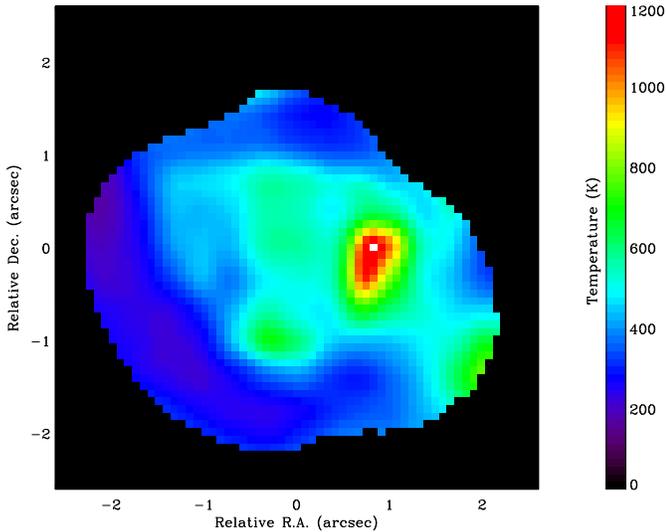}
  \caption{Color map of V\,Hya. Notice that the emission peak is
   off-centered what is probably due to the presence of a jet roughly oriented
   in the east-west direction.}
  \label{color_temp_vhya}
  \end{figure}

Using the method proposed by Dayal et al. (1998), we derived a color 
map (9.78 and 11.65\,$\mu$m filters) for V\,Hya, which gives an indication of the 
mean dust temperature along the line of sight,
weighted by the density distribution (Fig. \ref{color_temp_vhya}). The
meaning of such an average temperature is discussed in some more detail in Section\,4.

\begin{figure}
 \includegraphics[width=8.5cm]{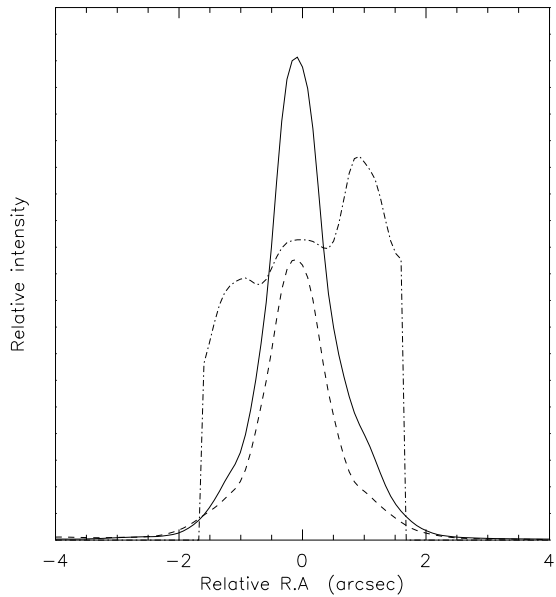}
 \caption{Cut in the color map of V\,Hya along the direction defined by the
center of the envelope and the color maximum (dot-dashed curve). The solid
and the dashed curves are cuts in the same direction made in the N2 and SiC
filter maps respectively.}
 \label{cutVhya}
 \end{figure}

One striking aspect of the color map is that the maximum
is off-centered. If we perform a cut in the color map along the direction defined by the
center of the envelope (the peak intensity coincides in both filters)
and the color maximum (Fig. \ref{cutVhya}), we observe three distinct 
peaks: a fainter one at $0.9\,\arcsec$ east from the center; a slightly higher peak
at the center of the envelope and a still higher peak at $0.9\,\arcsec$ west from the
center. It is worth mentioning that the direction of this cut is approximately
the same as that of the jet observed by Sahai et al. (2003) through the [S~II] emission
as well as inferred from interferometric CO (J=1-0) maps. Notice that the epoch
of these observations correspond to 2002 and 1998 respectively.
The jet observed by Sahai et al. (2003) has a characteristic expansion timescale 
of $\sim\,3$ years and an opening angle of $\sim\,20^o$\,-\,$\,30^o$\,sin\,$i$. The estimated 
timescale suggests that the events observed in 1998 and 2002 are probably not the 
same. Thus, they suggested that V\,Hya undergoes periodic mass ejection events
along a direction perpendicular to the equatorial disk.
Our color map corresponds to data obtained in 1995, prior to the aforementioned
events. If we assume an expansion velocity of $\sim\,200\,\mathrm{km\,s}^{-1}$ and a distance
of 340\,pc, we find an expansion timescale
of $\sim\,8\,$yr (the peaks we observe are offset by 0.9\,\arcsec\, from the center and have
an opening angle of roughly $40^o$\,sin\,$i$). This timescale suggests that we are observing
a material ejected in a previous mass-loss episode and it is also consistent
with the 12\,yr period over which the high-velocity [S\,II] emission has been observed
and monitored from 1989 up to 2000 (Lloyd Evans 2000).

\subsection{Roberts\,22}
Roberts\,22 (IRAS\,10197-5750) is an optical ($10\,\arcsec\,\times24\,\arcsec$)
bipolar proto-planetary nebula (PPN), whose circumstellar dust envelope 
absorbs totally  the
light emitted by its central star (Allen et al. 1980). This object is also an 
OH maser source
displaying strong emission at 1612 and 1665\,MHz, indicating that the
circumstellar envelope expands with a velocity of about
$20\,\mathrm{km\,s}^{-1}$. Optical spectroscopy shows that the
$\mathrm{H}\alpha$ line 
profile has wings extending up to $\pm\,450\,\mathrm{km\,s}^{-1}$, a signature
of high-speed outflow. The bipolar shape of Roberts\,22 is confirmed by
HST observations. These observations show also a dust waist oriented 
along P.A. of 121$^o$, which darkens the central star and
separates the polar lobes (Sahai et al. 1999). It is interesting to note that Roberts\,22
display thin shell structures as seen in many PPN (e.g. CRL\,2688), which could be due to
discrete mass-loss episodes; In addition, a striking aspect of Roberts\,22 is that C-rich
and O-rich spectral features are seen as in  the case of Red
Rectangle nebula (Waters et al. 1998). A possible explanation is that the
O-rich material is a remnant of an earlier mass-loss event that occurred
when the envelope of the star was \- O-rich, followed by the capture and
formation of a disk-like structure around a companion, the C-rich
material  being ejected in a late evolutionary phase, when the star
atmosphere became C-rich.
 
\begin{figure}
   \centering
   \includegraphics[width=8.cm]{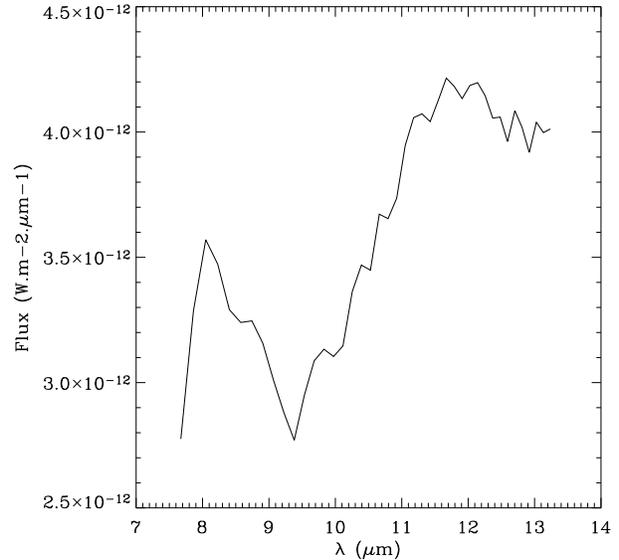}
   \caption{IRAS low resolution spectra of Roberts\,22.}
    \label{lrs}
    \end{figure}

\begin{figure*}
 \centering
 \includegraphics[width=18.cm]{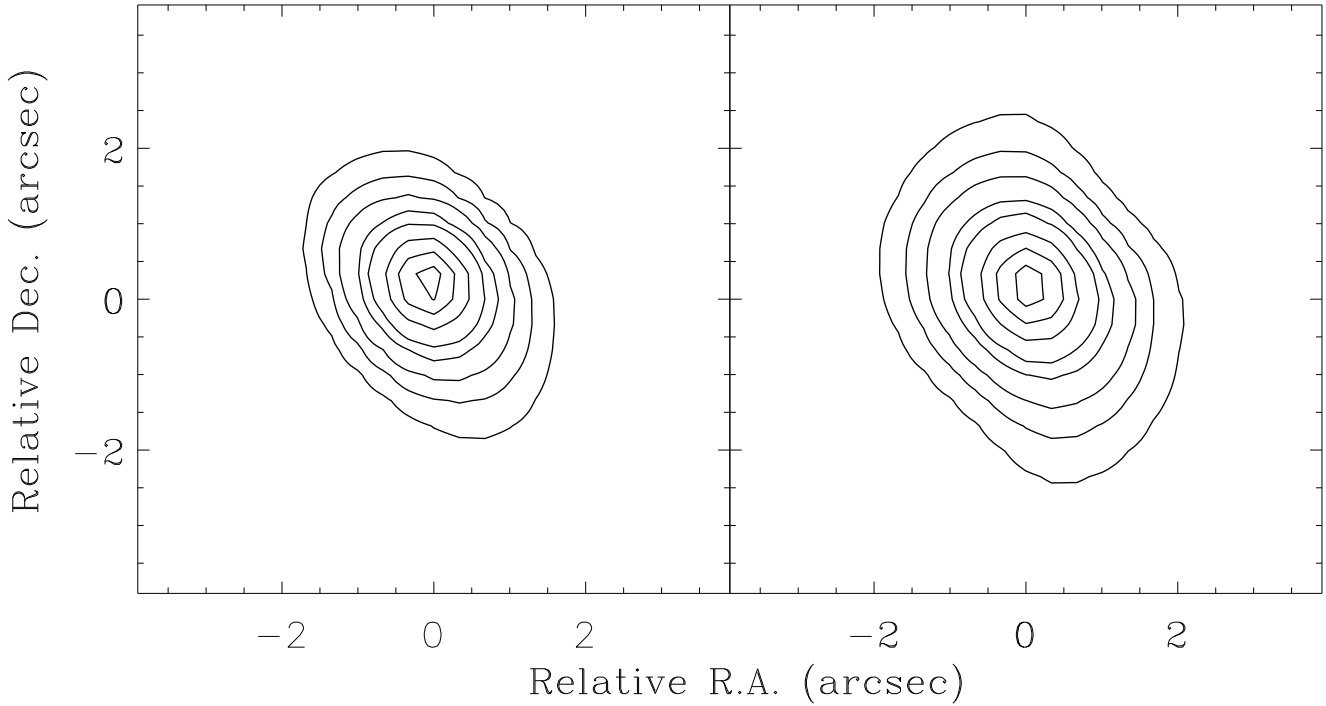}
   \caption{Contour plots of the 8.4\,$\mu\mathrm{m}$ (left) and 
   11.65\,$\mu\mathrm{m}$ (right) images of Roberts\,22. North
   is up and east is to the left. Contour levels are 90, 70, 50, 30, 20,
   10, 5 and 2\,\% of the peak surface brightness. }
  \label{FigRoberts22}
  \end{figure*}

Our 8.39\,$\mu \mathrm{m}$ and 11.65\,$\mu \mathrm{m}$ images of
Roberts\,22 show an envelope slightly elongated  in the direction north-east/south-west
with P.A. $\sim\,45^o$ (Fig. \ref{FigRoberts22}).  The SiC image (11.65\,$\mu$m)
resembles pictures obtained by the HST, since the X-shaped structure can be
guessed,  but being broader than in HST images.  
This structure could be due to the presence of an extended dust envelope
 not only along the equator but also inside the lobes observed in
 optical diffused light.
  Differently from V\,Hya and IRC\,+10216, IRAS low resolution
spectra of Roberts\,22 (Fig. \ref{lrs}) shows clearly two emission components, peaking
respectively around 8\,$\mu$m and 12\,$\mu$m. Our images were obtained
at these wavelengths, excluding the application of our method (see Section\,4) to
derive the distributions of density and temperature.

\begin{figure}
   \centering
   \includegraphics[width=9.cm]{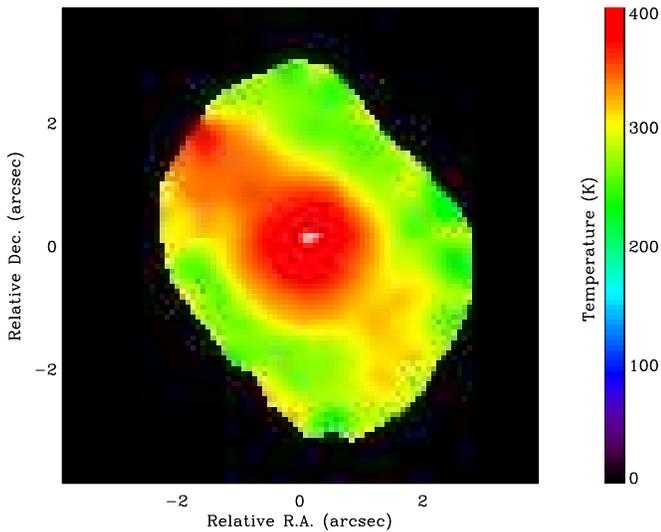}
   \caption{Color map of Roberts\,22.  Warm grains are observed at the
   center of the nebula, as well as in a jet-like structure 
   with P.A. $\sim\,45^o$.}
   \label{color_temp_r22}
   \end{figure}

 In Figure \ref{color_temp_r22}  a color map (8.39 and 11.65\,$\mu$m
filters) of Roberts\,22 is shown. In
this map the elongated north-east/south-west structure is clearly seen as well as 
some structures
suggesting the existence of local temperature variations, probably produced 
by different physical properties of dust grains inside the envelope.

\subsection{IRC\,+10216}
IRC\,+10216 (IRAS\,09452+1330) is the closest, and thus one of the most studied C-rich AGB
stars. Its circumstellar envelope is roughly spherical but composed of  many discrete
and incomplete shells  (Mauron \& Huggins 2000). From near-infrared speckle-masking
interferometry observations, Weigelt et al. (1998) were able to resolve five
individual clumps within a distance of 0.21\,\arcsec\, from the central
star. After six years of monitoring these structures, the onset of a fast wind phase
is hardly suggested  (Weigelt et al. 2002), although significant velocities might be guessed
through proper motion measurements.  These clumpy
structures seems to be consistent with the expectations of the 
Interacting Stellar Winds model (ISW) of Kwok et al. (1978), which describes 
correctly the shape of a planetary nebula in which a fast wind starting at the
end of the AGB phase interacts with a low velocity  fossil material.
IRC\,+10216 would presently be in the fast wind phase whose onset is relatively recent.

\begin{figure*}
  \centering
  \includegraphics[width=18.cm]{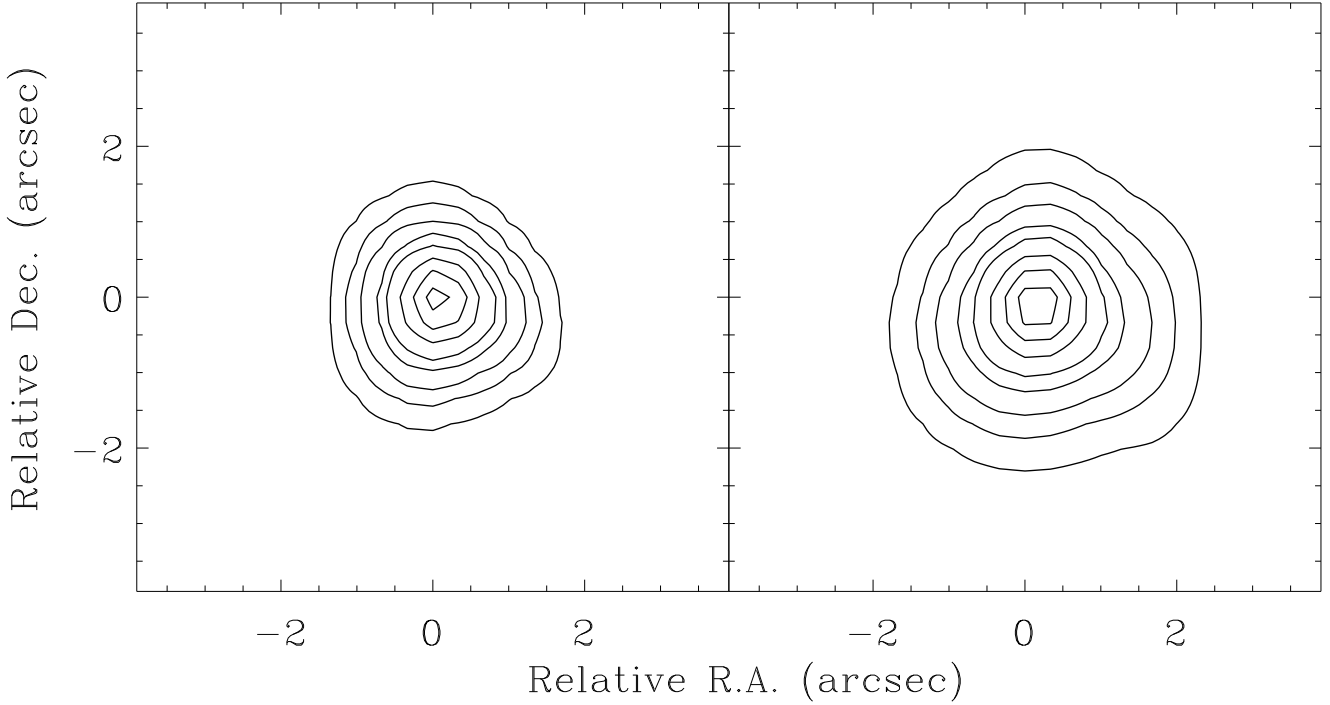}
   \caption{Contour plots of the 8.4\,$\mu\mathrm{m}$ (left) and 
   11.65\,$\mu\mathrm{m}$ (right) images of IRC\,+10216. North
   is up and east is to the left. Contour levels are 90, 70, 50, 30, 20,
   10, 5 and 2\,\% of the peak surface brightness. }
   \label{FigIRC+10216}
  \end{figure*}

A large number of observations of this star is available in the UBV
filters (Mauron et al. 2003), radio (Gensheimer et al. 1995) and
infrared (Weigelt et al. 1998 and references therein). However, to our 
knowledge,  no direct mid-infrared images have been published in the  literature. 
Here we report the first images of this star  in narrow band filters at
$8.39$ and $11.65\,\mu\mathrm{m}$ (N1 and SiC filters). In these images,
IRC\,+10216 appears to be roughly spherically symmetric, having a
spatial extensions at 2\,$\%$ intensity level of $3.5\,\arcsec\,\times\,3.5\,\arcsec$ 
and $4.5\,\arcsec\,\times\,4.5\,\arcsec$ in the N1 
and SiC filters, respectively (Fig. \ref{FigIRC+10216}).

\begin{figure}
   \centering
   \includegraphics[width=9.cm]{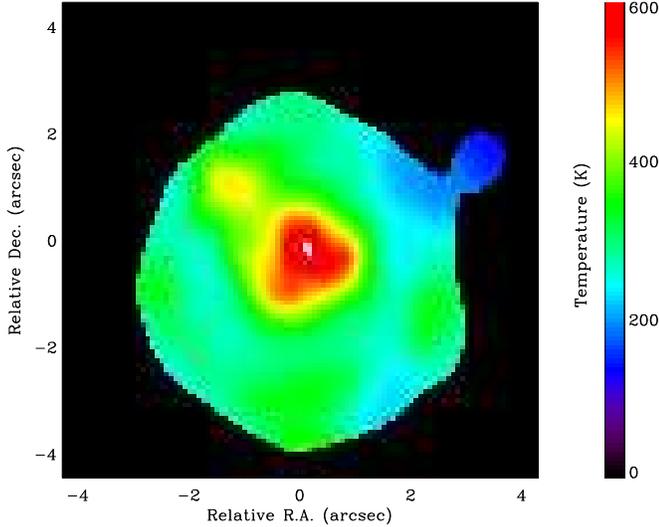}
   \caption{Color map of IRC\,+10216. The envelope is roughly
   spherically symmetrical but displays inhomogeneities in the dust 
   temperature distribution.}
   \label{color_temp_10216}
    \end{figure}

In Figure \ref{color_temp_10216} the color map (8.39 and 11.65\,$\mu$m filters) is
shown.  An elongated structure with the same orientation as the
near-infrared image of Weigelt et al. (1998) is clearly seen as well as inhomogeneities in the
temperature distribution. The cold region observed in the north-west is
probably an artifact since the signal-to-noise ratio is small at these intensity
levels.

\subsection{CIT\,6}
CIT\,6 (IRAS\,10131+3049) is an ``extreme'' carbon star believed to be
undergoing a transition  between the AGB phase to the PN stage
(Trammell et al. 1994).  CIT\,6 is the second brightest carbon star 
at 12\,$\mu$m after IRC\,+10216. It is a long-period-variable star with a period 
of $\sim\,628$ days (e.g. Alksnis 1995), and was classified as PPN by Johnson \& Jones
(1991). However, the estimated photospheric temperature
of $\sim$\,2\,800\,K (Cohen 1979) implies that the star is still near the
AGB. Visible and infrared polarization measurements show that the light at these
wavelengths is highly polarized, with a strong variation of the
polarization angle with the wavelength, indicating that the
distribution of the circumstellar material around
this star is asymmetric (Trammell et al. 1994 and references therein).   

 CIT\,6 has been observed  with a $\sim$\,50\,mas spatial resolution 
at 2.2\,$\mu$m and 3.1\,$\mu$m, using aperture masking techniques at the 
Keck\,I Telescope (Monnier et al. 2000). The object was resolved into
two main components: a bright elongated feature showing shape 
variations (north component) and a fainter bluer (south) component 
about 66\,mas away. The observed shape variations suggest an evolution of the
dust envelope, which can be explained by changes in the illumination pattern. 
CIT\,6 was also observed with WFPC2 aboard the
HST,  through the filters F439W\,(0.429\,$\mu$m), F555W\,(0.525\,$\mu$m), and 
F675W\,(0.674\,$\mu$m). Both components are seen in F675W and F555W images
 while in the F439W image only the south component is seen. These observations
are consistent with a binary star scenario where the north component presents
a 640 day pulsation period not seen in the south component, probably a main
sequence A-F star. Associated to the latter, a cometary-like feature extending along
the south-west direction can be seen in the 3.1\,$\mu$m images (Monnier et al. 2000).  
Recently, Schmidt et al. (2002) concluded from HST imaging,
ground-based spectroscopy and spectropolarimetric observations,
that a third closely coupled companion would be required to 
explain the bipolar morphology of this star. 

\begin{figure*}
 \centering
 \includegraphics[width=18.cm]{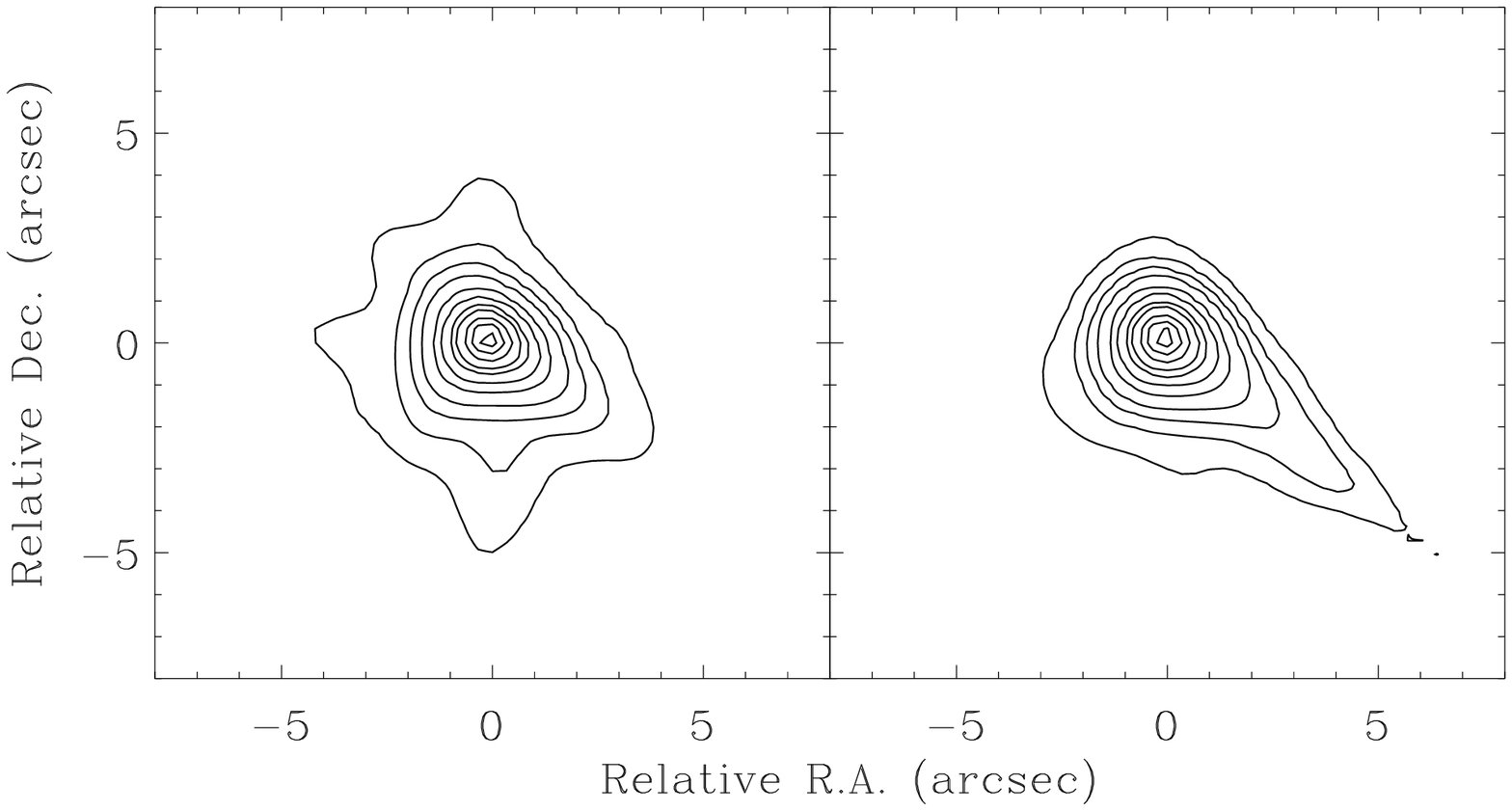}
 \caption{Contour plots of the 8.4\,$\mu\mathrm{m}$ (left) and 
   9.7\,$\mu\mathrm{m}$ (right) images for CIT\,6. North
   is up and east is to the left. Contour levels are 90, 70, 50, 30 , 20,
   10, 5 , 3, 2 and 1\,\% of the peak surface brightness.}
  \label{FigCit6}
  \end{figure*}

Here we report the first mid-infrared imaging  of 
CIT\,6 at 8.39\,$\mu\mathrm{m}$\,(N1) and  9.7\,$\mu\mathrm{m}$\,(Silicates).
The cometary-like feature is clearly seen in the 9.7\,$\mu$m image
(Fig. \ref{FigCit6}) extending up to 7\,$\arcsec$ (at 1\,$\%$ intensity
level) from the center of the
envelope at P.A. $\sim\,230^o$ (the north and south components seen in the near infrared 
are not resolved in our images). This cometary-like feature which is less
extended and less prominent in the 8.4\,$\mu$m image compares well
with pictures at 3.1\,$\mu$m (Monnier et al. 2000) and is roughly
orthogonal to the extended structure seen in the F439W HST image.
The morphology of CIT\,6 seen in the optical and near-infrared domains
can qualitatively be explained if  the bipolar 
nebula is essentially seen through its scattered light. The central star is
obscured by a central disk or torus of dust. The dominance of the
northern pole observed in the near-infrared indicates an inclination
toward us (Monnier et al. 2000).  This picture is supported by 
linear polarization data in the infrared but it does not account for the 
cometary shape seen in blue WFPC2 images (Trammell et al. 1994) and
in our mid-infrared images. No satisfactory flux calibration of our images
was obtained.

\section{The Model}
In spite of the presence of some asymmetries in the outer envelope of these 
objects, here we consider a spherical geometry as a  first approximation to
derive the temperature and  the density profiles of the circumstellar
dust around V\,Hya and IRC\,+10216. Roberts\,22 and CIT\,6 were excluded
from our modeling, since the former has at least two emission components and the
latter shows clearly elongated  isophotes. Our 
procedure is the following: in a first step, from the isophotal contours  we derive
the equivalent radius $a$ defined by  the relation $a = \sqrt{({S}/{\pi}})$, where
$S$ is the area delimited by the isophote of intensity $I(a)$.

   \begin{figure}
   \centering
   \includegraphics[width=6.cm]{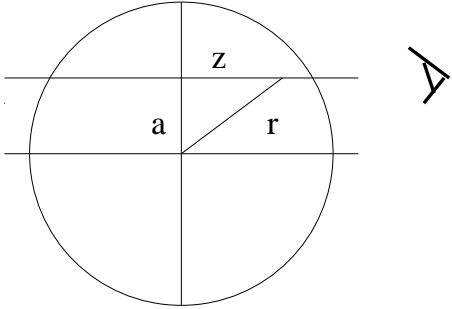}
   \caption{Schematic view of the nebula considered as spherically
   symmetric. The observer is on the right. So we have {\bf$I(a)=2\int_{0}^{z
   max}j(r) dz$}, where \textit{I} is the intensity 
   received and \textit{j} the emissivity of the dust grains. }
   \label{Fig_model}
   \end{figure}

If the outer nebular light at $\sim$\,10\,$\mu$m 
is dominated by the emission of warm dust grains and if the nebula 
is optically thin at  these wavelengths, then the intensity at a projected distance
$a$ from the center is

\begin{equation}\label{}
I(a) = 2\int^{R_e}_a j(r)\frac{rdr}{\sqrt{r^2-a^2}}
\end{equation}
where $R_e$ is the outer radius of the dust envelope and $j(r)$ is the dust 
emission coefficient, which is a function of  the
distance $r$ from the star. 
The equation above can easily be written under the form of an Abel integral, whose
inversion method is quite well known. Therefore, the dust emission coefficient is given by

\begin{equation}\label{}
j(r) = -\frac{1}{\pi}\int^{\infty}_r \frac{\frac{\partial I}{\partial a}da}{\sqrt{a^2-r^2}}
\end{equation}

On the other hand, from its definition, the dust emissivity can explicitly be written as

\begin{equation}\label{}
j(r) = n_g<r_g^2>Q_{\lambda}B_{\lambda}(T_g)
\end{equation}
where $n_g$ is the dust grain density, $<r_g^2>$ is the dust grain mean square
radius, $Q_{\lambda}$ is the emission (absorption) efficiency and $B_{\lambda}(T_g)$
is the Planck's function at the grain temperature $T_g$. If the dust
emissivity is derived at two different wavelengths $\lambda_1$
and $\lambda_2$, then their ratio is simply
\begin{equation}\label{}
\frac{j_{\lambda_1}}{j_{\lambda_2}} = \frac{Q_{\lambda_1}}{Q_{\lambda_2}}
\frac{B_{\lambda_l}(T_g)}{B_{\lambda_2}(T_g)}
\end{equation}

The grain temperature can be derived from Eqs.(2) and (4) if we assume a given
chemical composition and dimension for the dust grains. Once the
temperature distribution is obtained, the grain density can be evaluated from Eq.(3).
Notice that larger is the difference between $\lambda_1$ and $\lambda_2$, higher is the
accuracy in the determination of the temperature distribution and the
grain density. 
 
The emission efficiency of the grains $Q_{\lambda}$  was computed using the Mie's theory,
 assuming homogeneous and spherical grains constituted of amorphous carbon with  a mass
density of $2 \, {\mathrm {g \, cm}}^{-3}$. Optical properties were taken
from laboratory data (Rouleau \& Martin 1991; Suh 2000; Jager et al. 1998, hereafter
respectively Model 1, 2 and 3).  The choice of  dust  parameters is somewhat
arbitrary, since there are no strong observational constraints on them. Thus, grain
dimensions were chosen to be comparable to those adopted in previous studies,
 whenever existent.

For numerical reasons, in order to invert the Abel integral, we preferred to
adopt an analytical representation for the intensity $I(a)$ derived from our data.
In fact, the function
\begin{equation}\label{}
 I(a) = \frac{A}{\lbrack 1 + (\frac{a}{r_0})^{\gamma}\rbrack}
\end{equation}
represents quite well the observed intensity profiles. The resulting
fit parameters for the different objects are shown in
table (\ref{table_param}).

To verify the consistency of our assumption, namely, an
optically thin envelope at the considered wavelength, the optical depth at 
11.6$\mu$m  was computed from the equation below and the derived
density distribution.

\begin{equation}\label{}
\tau =\int \pi r_g^2Q_{\lambda}n_g(r)dr,
\end{equation}

Grains sizes ranging from 0.007\,$\mu$m up to 5\,$\mu$m
and dust optical properties taken from Models 1, 2 and 3 were used.
The derived optical depth values  (Table \ref{tabletau})
 are consistent with the hypothesis of an envelope optically thin
at $\sim$\,10\,$\mu$m, since for all models $\tau \leq 0.3$.
As an additional check, other flux values in the interval 8\,-\,12\,$\mu$m 
were computed  and compared with IRAS data. The good agreement between
expected and observed values reinforce the basic assumption of our model.

Color maps are usually interpreted as "temperature maps" (Dayal et
al. 1998). If the medium is optically thin, the intensity at a given
wavelength is given by:

\begin{equation}\label{}
I_{\lambda}= \int{n_g<r_g^2>Q_{\lambda}B_{\lambda}(T_g)dz}
\end{equation}

Let us define:
\begin{equation}\label{}
\overline{Q_{\lambda}B_{\lambda}(T)}=\frac{\int{n_g<r_g^2>Q_{\lambda}
B_{\lambda}(T_g)dz}}{\int{n_g<r_g^2>dz}}
\end{equation}

Then, from Eqs. (7) and (8) one obtains:

\begin{equation}\label{}
\frac{I_{\lambda1}}{I_{\lambda2}}=\frac{\overline{Q_{\lambda1}
B_{\lambda1}(T)}}{{\overline{Q_{\lambda2}B_{\lambda2}(T)}}}
\end{equation}

Eq. 9 tells that a given color corresponds to a given temperature only if this 
quantity is constant along the line of sight or, in other words, if the temperature is uniform 
throughout the nebula. Our model implies the presence of temperature
gradients, which are clearly seen as, for instance, in the color
maps of Roberts\,22 and IRC\,+10216. However color inhomogeneities are also detected, which 
should be interpreted as temperature variations along the line of
sight. These local variations are 
probably due to non-uniformities in the grain properties, which affect
the equilibrium temperature and the grain emissivity (see Eq.\,9).
Therefore the color maps should only be considered as a temperature "indicator".

\begin{table*}
\begin{center}
 \caption[]{Best fit parameters appearing in Eq.\,(5).}
 \label{table_param}
  \begin{tabular}[]{l l c c c }
  \hline
  \hline
  \noalign{\smallskip}
   Object & Filter 
   & A (${\mathrm {erg}}~{\mathrm {cm}}^{-2}~{\mathrm s}^{-1}~{\mathrm
  {arcsec}}^{-2}~{\mathrm {Hz}}^{-1}$)  & r$_o$ (arcsec)  & $\gamma $  \\ 
           \noalign{\smallskip}
            \hline
            \noalign{\smallskip}
 V\,Hya       & SiC                  & 4.86\,$\pm\,0.05$\,$\times$10$^{-21} $  
              & 0.49\,$\pm\,0.05 $   & 2.66\,$\pm\,0.05$  \\
              & N2                   & 4.79\,$\pm\,0.05$\,$\times$10$^{-21} $  
              & 0.48\,$\pm\,0.05 $   & 2.70\,$\pm\,0.05$  \\
 IRC\,+10216  & SiC                  & 1.03\,$\pm\,0.01$\,$\times$10$^{-19} $  
              & 0.73\,$\pm\,0.01 $   & 2.47\,$\pm\,0.05$  \\
              & N1                   & 1.24\,$\pm\,0.01$\,$\times$10$^{-19} $  
              & 0.63\,$\pm\,0.01 $   & 2.64\,$\pm\,0.07$  \\
           \noalign{\smallskip}
            \hline
         \end{tabular}
   \end{center}
   \end{table*}

\section{Results and discussion}

   \begin{figure*}
\includegraphics[width=8.75cm]{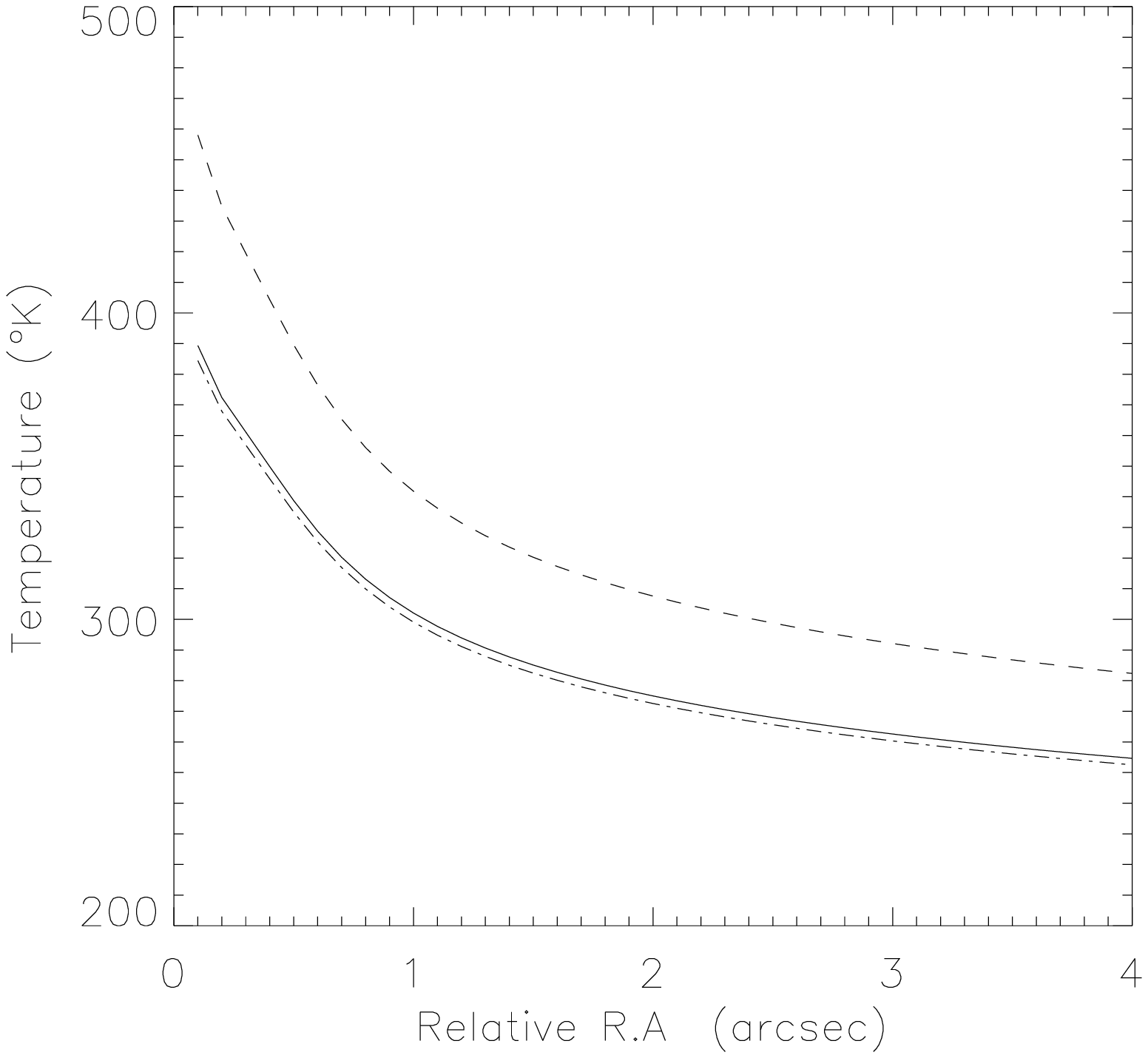}\includegraphics[width=8.75cm]{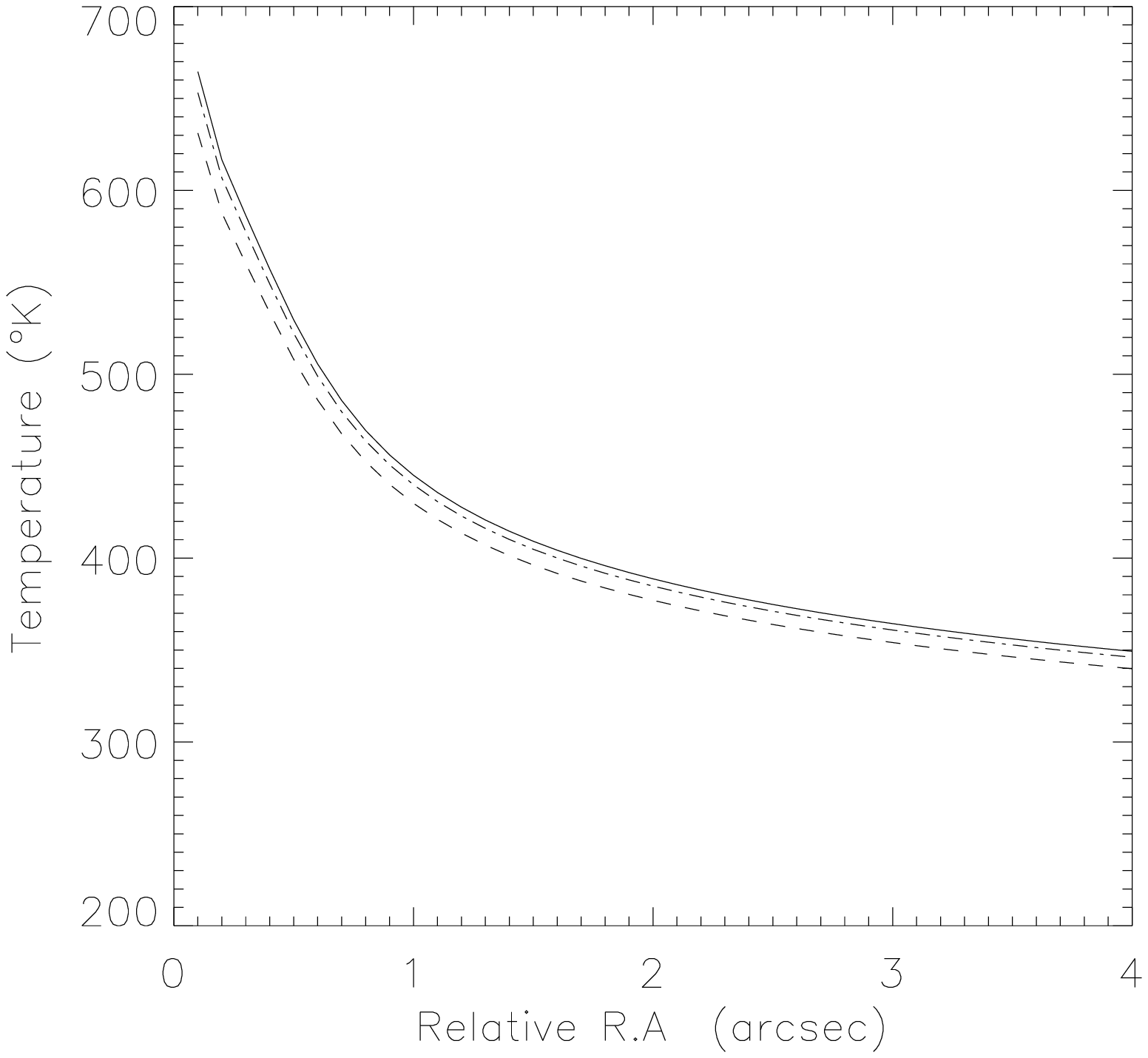}
\includegraphics[width=8.75cm]{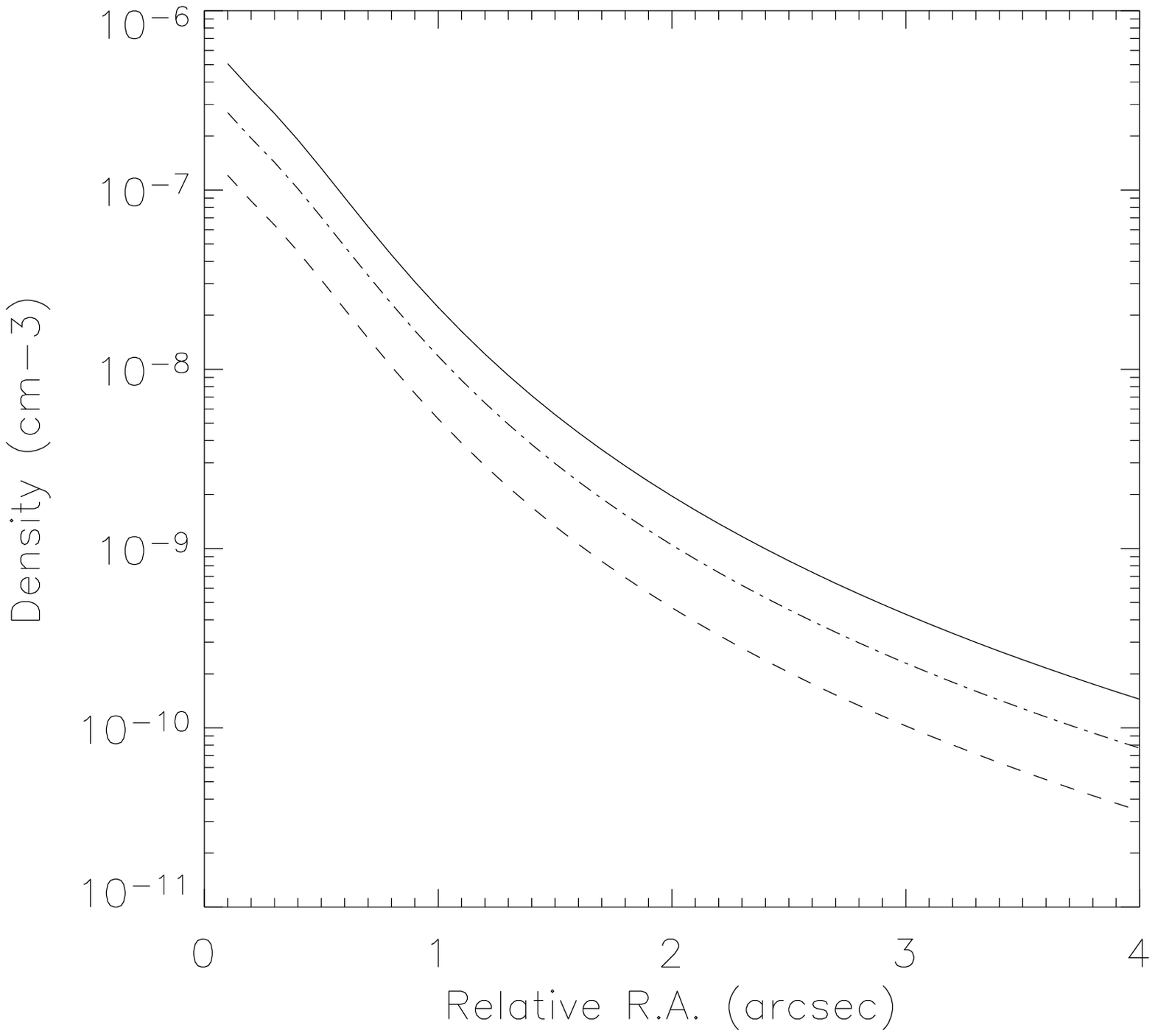}\includegraphics[width=8.75cm]{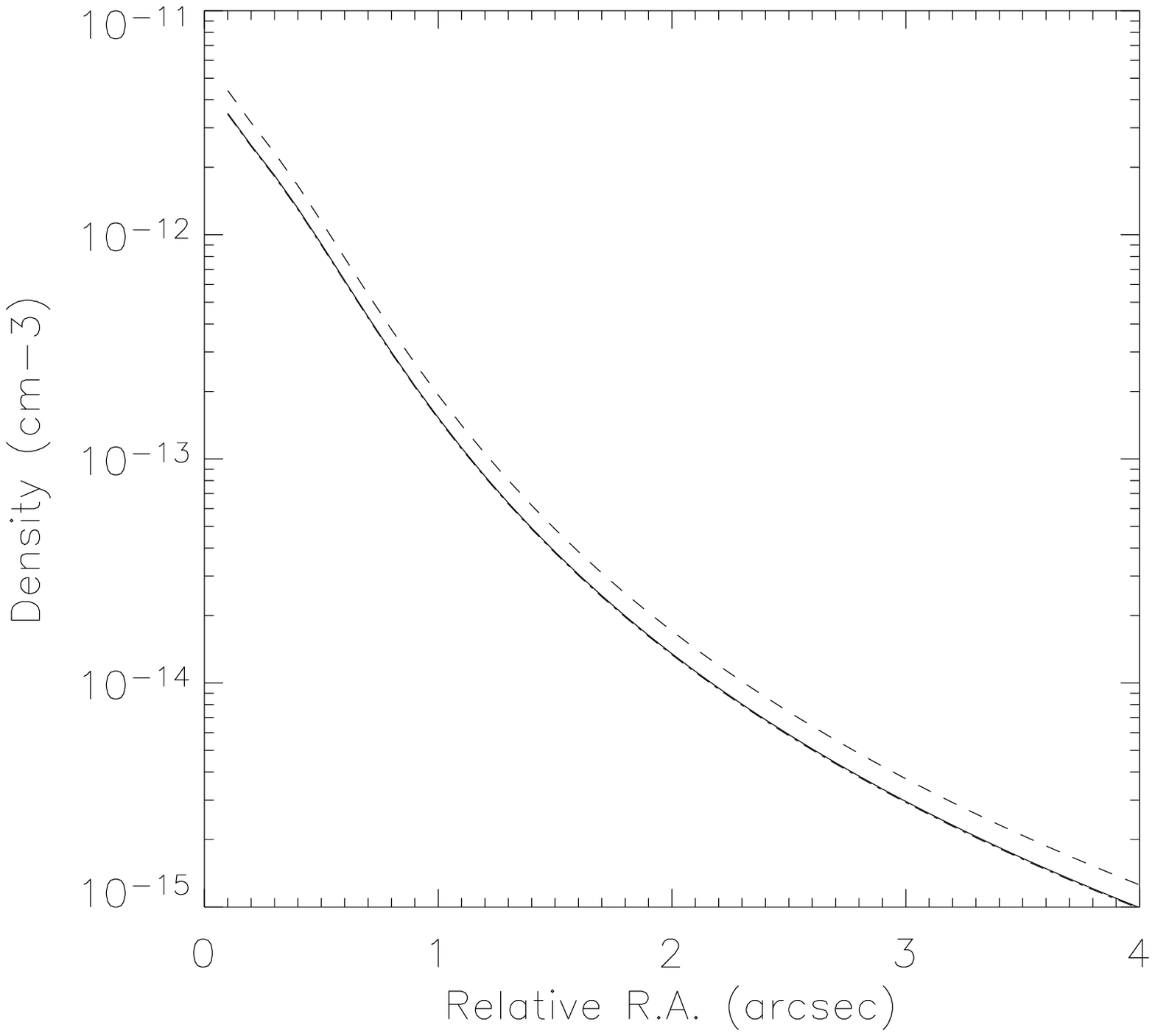}
\caption{Temperature and density profiles of V\,Hya obtained with the
 model parameters: d\,=\,340\,pc, $r_g$\,=\,0.2\,$\mu$m (left) 
and  d\,=\,340\,pc, $r_g$\,=\,5\,$\mu$m
 (right) respectively. The solid, dot-dashed and dashed  lines curves
 represent results from Models 1, 2 and 3 respectively.}
 \label{FigTempdensVhya}
 \end{figure*}

We have applied the aforementioned procedure to V\,Hya and
IRC\,+10216 in order to derive temperature and density
profiles for the dust inside the envelope of those stars. Once the density
profile is obtained, the total dust mass ($M_{\mathrm {dust}}$) can be calculated from 
\begin{equation}\label{}
M_{dust} = \int^{R_e}_{R_i} 4\pi r^2 n_g(r) m_g dr
\end{equation}
where $m_g={4\pi\delta  r_g^3}/{3}$ is the mass of a dust grain
($\delta$ being equal to $2.0\,{\mathrm {g\,cm}}^{-3}$ for
the amorphous carbon), $R_i$ and $R_e$ are  respectively the internal and external 
radius of the dust envelope. If we assume that the gas-to-dust mass
ratio $\xi$ and the expansion
velocity $V_e$ are constant throughout the dust envelope, then
an estimate of the mass-loss rate $\dot{M}$  can be obtained from
\begin{equation}\label{}
\dot{M} = \xi M_{dust}\frac{V_e}{R_e}. 
\end{equation}

Values for the gas-to-dust ratio $\xi$ adopted by different authors range
from 200  (Groenewegen 1997) up  to 360 (Knapp et al. 1997).
In the  present calculations,  the
value $\xi$\,=\,300 was adopted and for the expansion velocity, the values derived
from CO observations by Loup et al. (1993) were used.  
It is worth mentioning that whenever the radiation pressure drives
 the mass-loss,  dust velocities are known to be usually higher than 
 gas velocities, but differences are expected to be small if  drag forces 
are effective (Steffen et al. 1997).
 For the external radius $R_e$, we have adopted the values derived from our images.
As we shall see later, the dust density decreases approximately as $r^{-2}$ and, in 
this case, the mass-loss rate is almost independent on the value of $R_e$.

 \begin{table}
 \caption[]{Estimated dust optical depths  at $11.6\,\mu{\mathrm m}$ for 
 different grain sizes and optical properties. $\tau(1)$, $\tau(2)$ and $\tau(3)$ 
 correspond to Models 1, 2 and 3  respectively.}
 \label{tabletau}
 \begin{tabular}[]{l l c c c}
 \hline
 \hline
 \noalign{\smallskip}
  Object  & $r_g$  & $\tau$ (1) & $\tau$ (2) & $\tau$ (3)\\
 \noalign{\smallskip}
 \hline
 \noalign{\smallskip}
V\,Hya    & 0.2     & $2.8\times10^{-2}$ & $2.9\times10^{-2}$ 
          &$1.7\times10^{-2}$ \\
          & 5.0     & $7.5\times10^{-3}$ & $7.7\times10^{-3}$ 
          &$8.3\times10^{-3}$ \\
IRC\,+10216 & 0.007 & $7.9\times10^{-2}$ & $7.9\times10^{-2}$
          & $2.3\times10^{-2}$ \\
          & 0.008   & $7.9\times10^{-2}$ & $2.3\times10^{-2}$
          & $4.7\times10^{-2}$  \\
          & 0.015   & $2.3\times10^{-1}$ & $1.4\times10^{-1}$ 
          & $1.9\times10^{-1}$ \\
          & 0.05    & $2.3\times10^{-1}$ & $1.9\times10^{-1}$ 
          & $2.6\times10^{-1}$ \\      
          & 0.16    &$2.6\times10^{-1}$  & $2.2\times10^{-1}$
          & $2.7\times10^{-1}$  \\ 
          & 5.0     &$6.0\times10^{-2}$  & $6.6\times10^{-2}$
          & $6.8\times10^{-2}$  \\
 \noalign{\smallskip}
 \hline
 \end{tabular}
 \end{table}

\subsection{V\,Hya}
Some radiative transfer models have already been proposed to represent the circumstellar
envelope of this star. Kahane et al. (1996) have modeled this object
using millimeter CO observations and an axisymmetric model. 
Distance estimates for V\,Hya are in the range  400\,pc (derived
from  Hipparcos data) up to 550\,pc (derived by Bergeat et al. 1998
from the period-luminosity relationship for carbon stars). However,
lower values can be found in the literature. Kahane et al. (1996) adopted a distance
of 340\,pc and derived  a total mass-loss rate 
of $\sim$$\,1.5\,\times\,10^{-6}{\mathrm M}_{\sun}$/yr and a total mass
(gas+dust) for the circumstellar envelope 
of  $\sim$$\,2.1\,\times\,10^{-3}{\mathrm M}_{\sun}$. Knapp et al. (1997) 
modeled the dust envelope of V~Hya by assuming grains constituted of amorphous 
carbon with dimensions of $0.2\,\mu{\mathrm m}$  and by adopting a
distance of 380\,pc. They derived an envelope
 mass of $8.0\,\times\,10^{-3}{\mathrm M}_{\odot}$ and a mass-loss rate of
$2.5\,\times\,10^{-5}{\mathrm M}_{\odot}$/yr. These values depend on the
optical properties of the grains but are significantly higher than those obtained
by Kahane et al. (1996).

In our calculations, the distance to the star was assumed to be in 
the range 340\,-\,550\,pc. The masses of the dust envelope and the corresponding
mass-loss rates obtained  for different dust grain parameters and distances
are given in  Table \ref{table_mass} and in Table \ref{table_massloss}. Previous estimates 
 are also given. The temperature and density profiles of V\,Hya obtained with the
 parameters d\,=\,340\,pc; $r_g$\,=\,0.2\,$\mu$m  and  $r_g$\,=\,5\,$\mu$m are shown 
in Fig.\,\ref{FigTempdensVhya}. The temperature determination does not depend on
the assumed distance but depends on the adopted grain properties. 
For a grain size of 0.2\,$\mu$m, the dust temperature ranges
from 390\,-\,460\,K at the inner envelope down
to 255\,-\,270\,K at the outer parts of the envelope. The higher temperature values 
correspond to optical properties of Model 3. Higher temperatures,
ranging from 660\,K (inner envelope) down to 350\,K (outer
envelope) are obtained for larger grains ($r_g$\,=\,5\,$\mu$m) and, in this case,
no significant differences are seen for the different optical properties of
Models 1, 2 and 3.
 
If we compare our mass-loss rates derived with grain parameters 
similar ($r_g$\,=\,0.2\,$\mu$m) to those adopted by Kahane et al. (1996), we
notice that the results are comparable, in
spite of  differences in the methodology used by those authors with respect to our
approach.

\begin{table*}
\begin{center}
\caption[]{Mass of the dust envelope of V\,Hya and IRC\,+10216
for different optical models, distances (d) 
and  grain sizes ($r_g$) compared to values derived   
from the literature.  The inner and outer integration radii were
0.1\,\arcsec\, and 4\,\arcsec\, for V\,Hya,  0.1\,\arcsec\, 
and 10\,\arcsec\, for IRC\,+10216. Dust masses taken from the
literature have been scaled to correspond to the same outer radii as
those used in this present work. Expansion velocities have been taken  equal to  
$18.7\,{\mathrm {km\ s}}^{-1}$ and $14.7\,{\mathrm
{km\,s}}^{-1}$ for V\,Hya and IRC\,+10216 respectively.}
\label{table_mass}
 \begin{tabular}[]{l l l l c l | l  l}
 \hline
 \hline
 \cline{4-8}
 &  &   & & Present Models&   & ~~~~~~~~~~~~Previous Models&\\
 &  &   & & ${\mathrm M}_{\mathrm {dust}}$(${\mathrm M}_{\sun}$) &  &   &\\
 &  &   & &  &  &   &\\
 Object    & $r_g(\mu{\mathrm m})$ &  d(pc) & Model 1&
 Model 2& Model 3 & ${\mathrm M}_{\mathrm {dust}}$(${\mathrm M}_{\sun}$)
 &Ref. \\
\hline
   &   &   & &     &   &  \\
V\~Hya       & 0.2  & 340    & $2.3\,\times\,10^{-6}$  & $1.2\,\times\,10^{-6}$ 
            & $5.4\,\times\,10^{-7}$   &  $2.2\,\times\,10^{-6}$ & (a)$^{\mathrm{*}}$ \\
            & 0.2  & 380    & $2.8\,\times\,10^{-6}$  & $1.5\,\times\,10^{-6}$ 
            & $6.7\,\times\,10^{-7}$   & $9.5\,\times\,10^{-6}$  & (b) \\
            & 0.2  & 550    & $5.9\,\times\,10^{-6}$  & $3.2\,\times\,10^{-6}$ 
            & $1.4\,\times\,10^{-6}$   &                         &  \\
            & 5    & 340    & $2.4\,\times\,10^{-7}$  & $2.4\,\times\,10^{-7}$ 
            & $ 3.1\,\times\,10^{-7}$  &                         &  \\
            &   &   & &     &   &  \\
IRC\,+10216
            & 0.007  & 170   & $1.5\,\times\,10^{-5}$ & $7.3\,\times\,10^{-6}$  &
            $ 2.1\,\times\,10^{-5}$    & $2.7$\,-\,$4.3\,\times\,10^{-4}$  & (c) \\
            & 0.008  & 150   & $1.3\,\times\,10^{-5}$ & $1.9\,\times\,10^{-5}$ &
            $ 3.9\,\times\,10^{-5}$ 
            & $0.3$\,-\,$3.8\,\times\,10^{-4}$ & (d)  \\
            & 0.015  & 130   & $2.6\,\times\,10^{-5}$ & $8.1\,\times\,10^{-6}$ &
            $ 8.7\,\times\,10^{-5}$    & $2.2\,\times\,10^{-5}$   & (e) \\  
            & 0.05   & 200   & $6.8\,\times\,10^{-5}$ & $3.0\,\times\,10^{-5}$ &
            $ 3.2\,\times\,10^{-5}$  & &  \\
            & 0.16   & 200  & $7.9\,\times\,10^{-5}$&$3.3\,\times\,10^{-5}$ &
            $ 3.2\,\times\,10^{-5}$     & $0.8$\,-\,$6.7\,\times\,10^{-4}$& (f) \\
            & 5.0    & 130  &$3.0\,\times\,10^{-6}$ &$3.2\,\times\,10^{-6}$ &
            $3.9\,\times\,10^{-6}$  &  &  \\
            & 5.0   & 200  &$7.2\,\times\,10^{-6}$ &$7.5\,\times\,10^{-6}$ &
            $ 9.3\,\times\,10^{-6}$  &  &  \\
\hline
\end{tabular}
\end{center}
 References : a) Kahane et al. (1996); b) Knapp et al. (1997);  c)
 Winters et al. (1994); d) Danchi et al. (1994); 
 e) Men'shchikov et al. (2001); f) Groenewegen (1997).   
\begin{list}{}{}
\item[$^{\mathrm{*}}$] {Based on radio observations.  No grain model was
  used in these models, then estimation of the dust mass was done assuming a gas-to-dust
 mass ratio of 300.}
\end{list}
\end{table*}
 
   \begin{figure*}
\includegraphics[width=8.75cm]{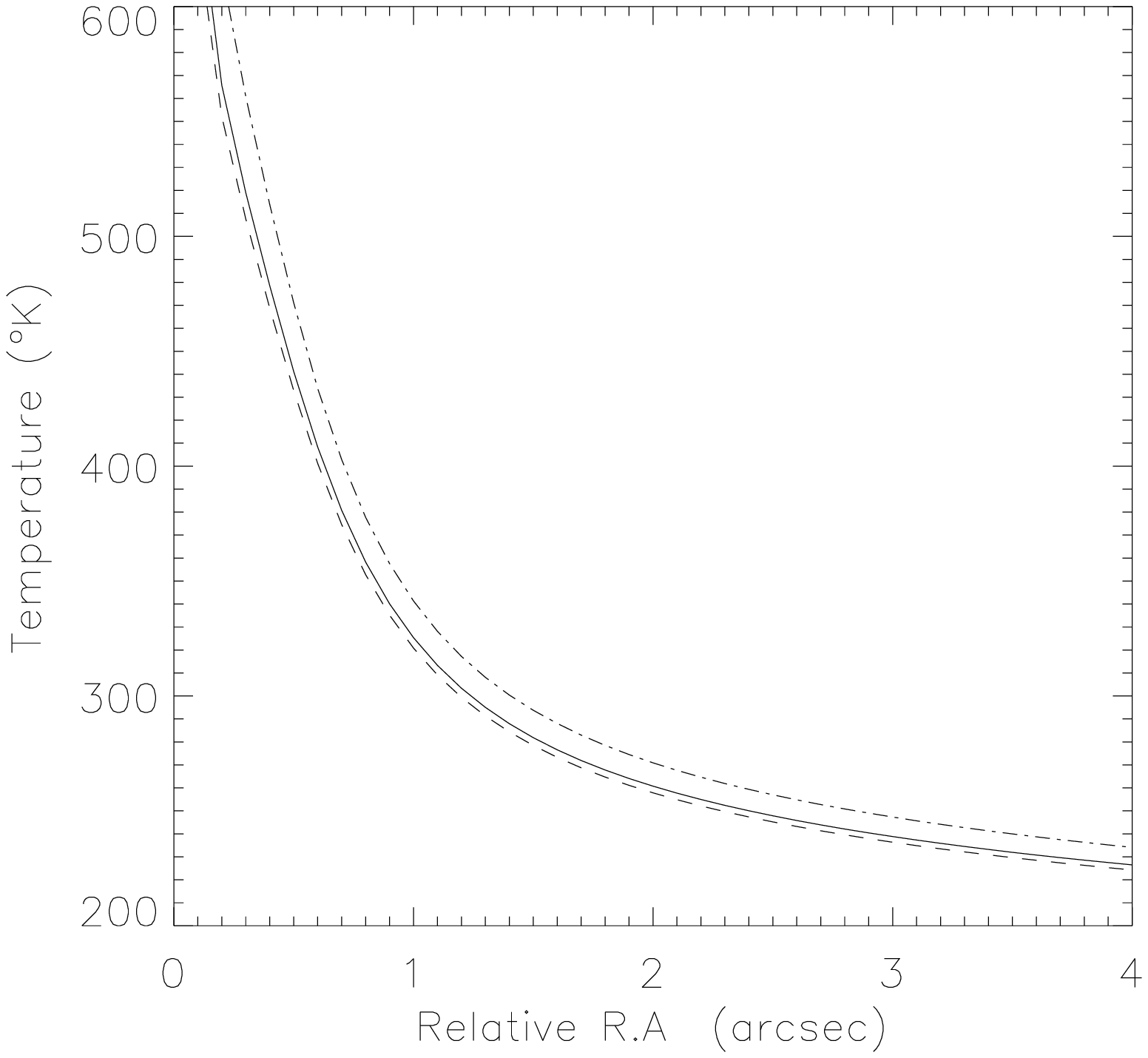}\includegraphics[width=8.75cm]{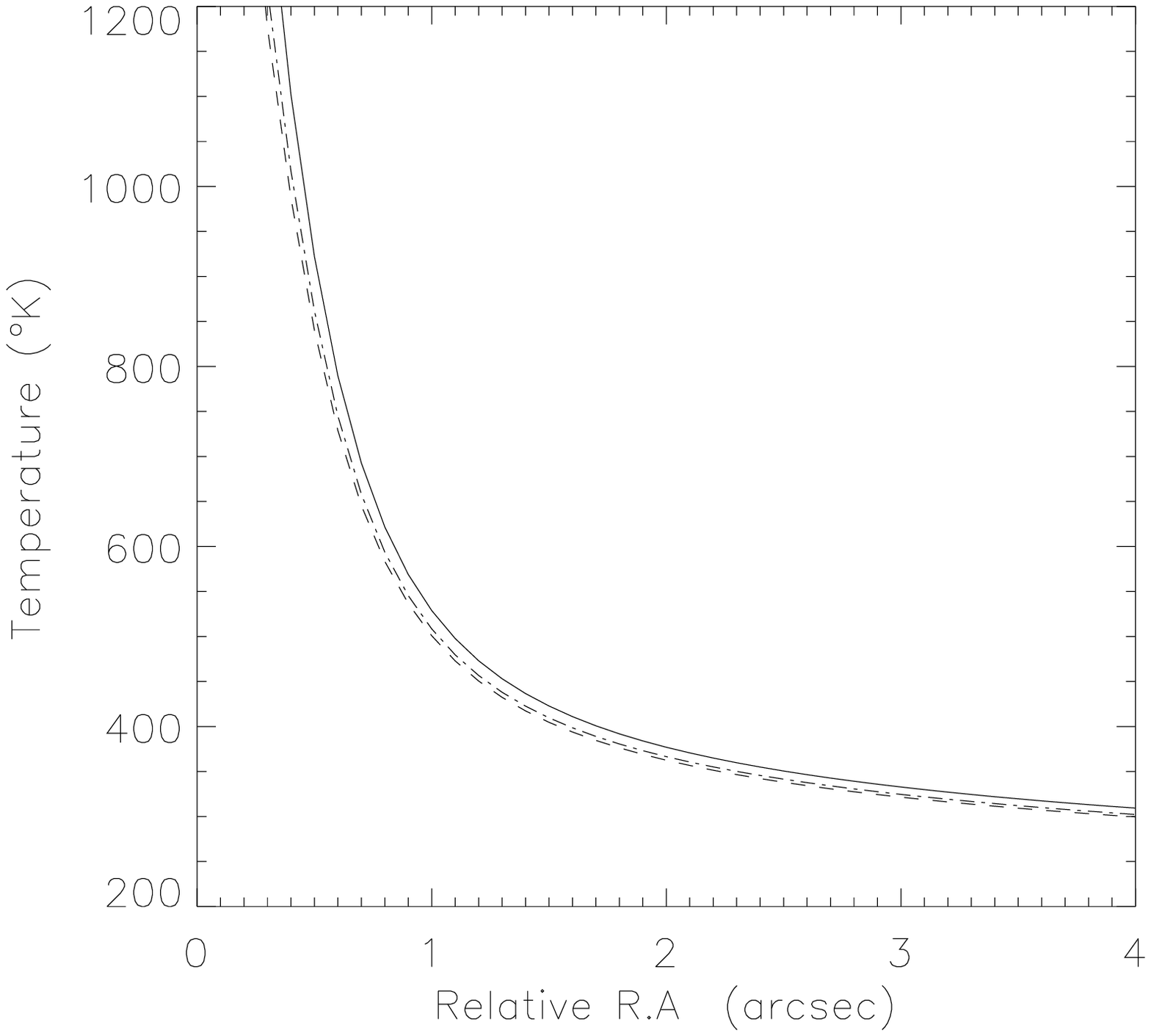}
\includegraphics[width=8.75cm]{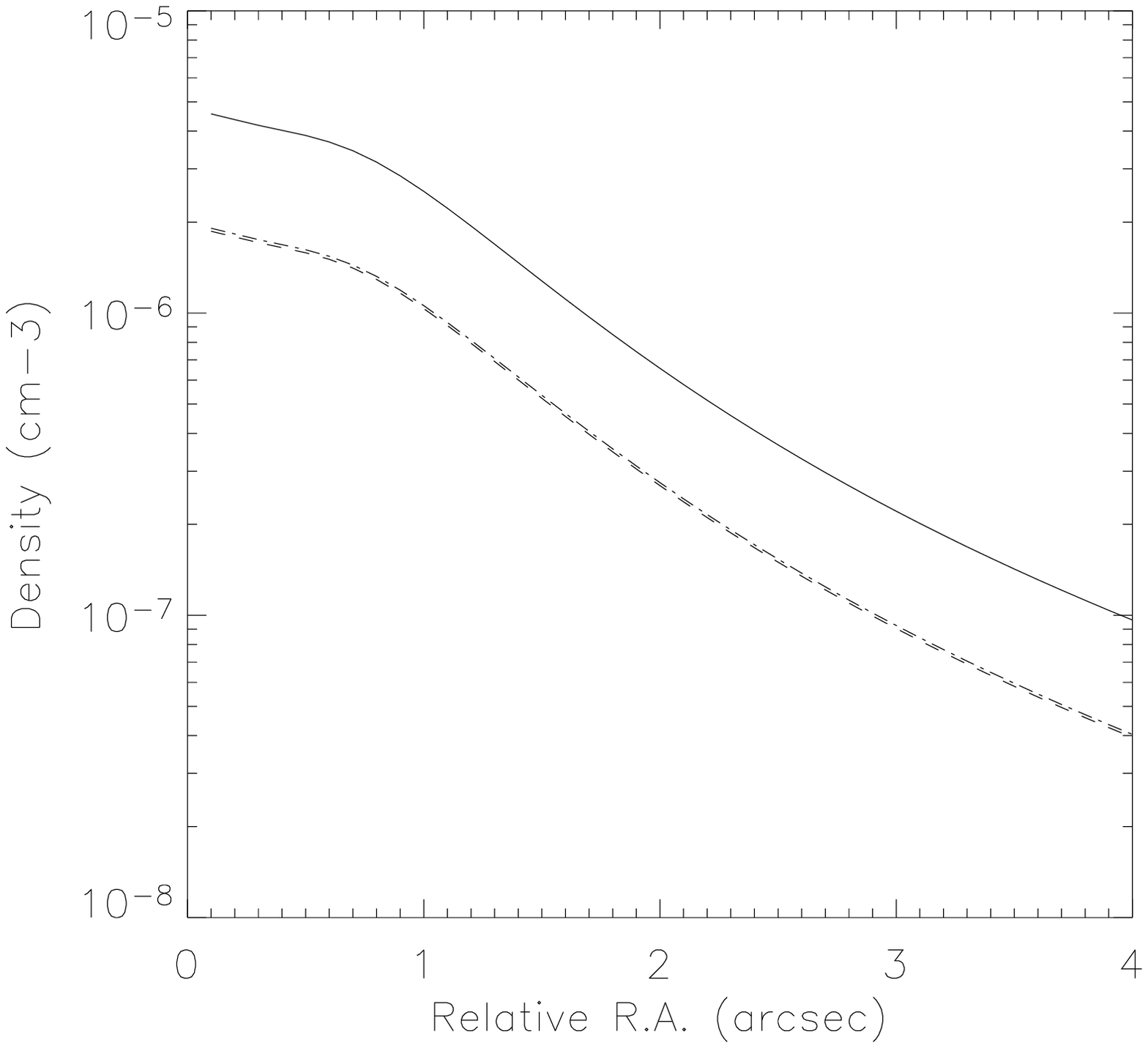}\includegraphics[width=8.75cm]{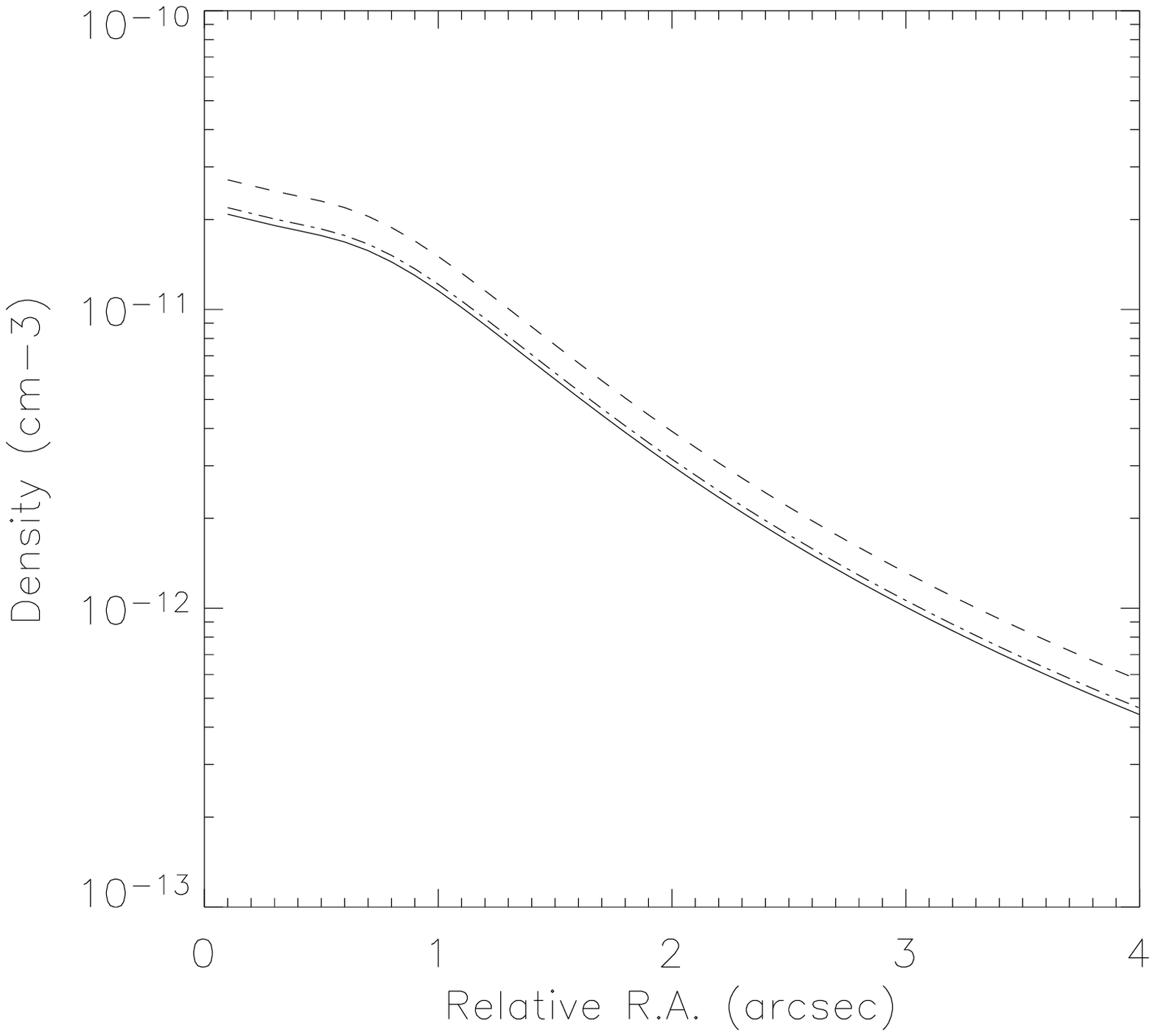}
 \caption{Temperature and density profiles of IRC\,+10216 obtained with the
 model parameters : d\,=\,200\,pc, $r_g$\,=\,0.16\,$\mu$m (left) and d\,=\,130\,pc,
 $r_g$\,=\,5\,$\mu$m (right) respectively. The solid, dot-dashed and dashed  lines curves
 represent results from Models 1, 2 and 3 respectively. }
  \label{FigTempdensIrc}
  \end{figure*}

\begin{table*}
\begin{center}
\caption[]{Mass-loss rates for V\,Hya and IRC\,+10216 derived from different parameters
and compared with previous determinations as in Table \ref{table_mass}.} 

 \label{table_massloss}
 \begin{tabular}[]{l l l l c l | l  l}
 \hline
 \hline
 \cline{4-8}
 &  &   & & Present Models&  &~~~~~~~~~~~~Previous Models &\\
 &  &   & & $\dot{\mathrm M}$ (${\mathrm M}_{\sun}{\mathrm {yr}}^{-1}$)&  &  &\\
 &  &   & & &  &  &\\
 Object    & $r_g(\mu{\mathrm m})$ &  d(pc) & Model 1&
 Model 2 & Model 3 & $\dot{\mathrm M}$ (${\mathrm M}_{\sun}{\mathrm {yr}}^{-1}$) &Ref. \\
\hline
   &   &   & &     &   &  \\
V\,Hya      &  0.2   & 340   & $2.0\,\times\,10^{-6}$ & $1.0\,\times\,10^{-6}$ 
            &  $4.7\,\times\,10^{-7}$    & $1.5\,\times\,10^{-6}$             & (a)\\
            & 0.2    & 380   & $2.2\,\times\,10^{-6}$ & $1.2\,\times\,10^{-6}$ &
            $ 5.2\,\times 10^{-7}$       & $2.5\,\times 10^{-5}$            & (b)  \\
            & 0.2    & 550   & $3.2\,\times\,10^{-6}$ & $1.7\,\times\,10^{-6}$ &
            $ 7.6\,\times 10^{-7}$       &             &  \\
            & 5     & 340    & $2.1\,\times\,10^{-7}$ & $2.1\,\times\,10^{-7}$ &
            $2.7\,\times\,10^{-7}$        &             &  \\
   &   &   & &     &   &  \\
IRC\,+10216
            & 0.007  & 170   & $8.0\,\times\,10^{-6}$ & $4.0\,\times\,10^{-6}$  &
            $1.1\,\times\,10^{-5}$     & $8.0\,\times\,10^{-5}$ & (c) \\
            & 0.008  & 150   & $8.1\,\times\,10^{-6}$ & $1.2\,\times\,10^{-5}$ &
            $2.4\,\times\,10^{-5}$     & $2.2$\,-\,$4.0\,\times\,10^{-5}$    & (d)  \\
            & 0.015  & 130   & $1.9\,\times\,10^{-5}$ & $5.8\,\times\,10^{-6}$ &
            $6.2\,\times\,10^{-6}$     & $1.6\,\times\,10^{-5}$  & (e) \\  
            & 0.05   & 200   & $3.2\,\times\,10^{-5}$ & $1.4\,\times\,10^{-5}$ &
            $1.5\,\times\,10^{-5}$     & $2.0\,\times\,10^{-5}$  & (g) \\
            & 0.16   & 200   & $3.7\,\times\,10^{-5}$ & $1.5\,\times 10^{-5}$ &
            $1.5\,\times\,10^{-5}$     & $4.9\,\times 10^{-5}$             & (f) \\
            & 5.0    & 130   & $2.2\,\times\,10^{-6}$ & $2.3\,\times\,10^{-6}$ &
            $2.8\,\times\,10^{-6}$   &             &  \\
            & 5.0    & 200   & $3.3\,\times\,10^{-6}$ & $3.5\,\times\,10^{-6}$ &
            $4.3\,\times\,10^{-6}$      &             &  \\
\hline
\end{tabular}
\end{center}
 References : a-f) as in Table \ref{table_mass}; g) Keady et
 al. (1988).
\end{table*}

\subsection{IRC\,+10216}
We have also applied our procedure to the dust envelope of IRC\,+10216 for
distances in the interval 130\,-\,200\,pc and dust grain sizes ranging from
0.007 up to 5.0\,$\mu$m. Our results are given in Tables\,\ref{table_mass},
\ref{table_massloss} and in Fig.\,\ref{FigTempdensIrc}.

For small grains (${\mathrm r}_g\,=\,0.16\,\mu$m), temperatures  attain
600\,K in the inner envelope and decrease to 220\,K in the outer parts. 
For larger grains (${\mathrm r}_g\,=\,5\,\mu$m)  temperatures in the inner region
are higher $\sim$\,1200\,K, decreasing to values around 300\,K in the outer regions of
the envelope. 

For a distance of 200\,pc, the derived dust masses are within the 
interval 0.7\,-\,7.9\,$\times\,10^{-5} {\mathrm M}_{\odot}$ and the
corresponding total (gas+dust) mass-loss rates are
$0.3\,-\,3.2\,\times\,10^{-5}\,{\mathrm M}_{\odot}$/yr.

Being one of the most studied carbon stars, several models for
the envelope of this object can be found in the literature.
Keady et al. (1988) modeled the envelope of this star using data
on the fundamental and first overtone lines of CO, assuming
a distance of 200\,pc and dust grains composed of amorphous carbon of size
$0.05\,\mu{\mathrm m}$. They found  temperatures
varying from $\sim$\,$1\,300$\,K at 13.5\,AU down to 300\,K
at $\sim$180\,AU. They have estimated a mass-loss rate close 
to $2\times\,10^{-5}{\mathrm M}_{\sun}$/yr. 
Groenewegen (1997) has considered the effects of a non-steady
mass-loss and estimated a total dust mass of about 
$1.0\,{\mathrm D}^2(\mathrm{kpc}){\mathrm M}_{\sun}$ for a variable
mass-loss rate,
which should be compared to $0.13\,{\mathrm D}^2({\mathrm {kpc}}){\mathrm M}_{\sun}$
derived under the usual assumption of a steady flow. An important aspect of this work 
is the constraint obtained for the grain sizes ($\sim\,0.16\,\mu$m),
derived from a large body of data. 
Winters et al. (1994) developed a quite sophisticated model, including 
a hydrodynamical gas-dust coupling, thermodynamics, radiative
transfer,  chemistry in the processes of  dust formation and growth. Assuming a
distance of 170\,pc and a mass-loss rate of $8\,\times\,10^{-5}{\mathrm
M}_{\sun}$/yr, they found a dust mass in the envelope within the
range 2.0\,-\,3.2\,$\times\,10^{-3}{\mathrm M}_{\sun}$. Similar dust masses
and mass-loss rates have been found by  Danchi et al. (1994) from mid-infrared data.
  
Most of the past studies suggest that  mass-loss rates for this object are in the range
 2\,-\,8\,$\times\,10^{-5}\,M_{\odot}$/yr if the distance is 170\,-\,200\,pc. The present 
analysis indicates that such values can only be obtained if grain dimensions are
in the interval 0.05\,-\,0.6\,$\mu$m.  Larger grains lead to mass-loss rates one
order of magnitude smaller than the aforementioned estimates.

\section{Conclusions}
In this paper, the first mid-infrared images of a sample of carbon-rich AGB stars
are reported. From our data, temperature and
density profiles of the dust envelope surrounding these stars were derived from a simple
inversion method applied to obtain the dust emissivity. 

Color maps are compatible with the presence of temperature gradients:
hot and/or warm dust grains are present in the central
regions whereas cold dust grains are seen in outer parts of the envelope. Moreover,
these color maps suggest the presence of temperature inhomogeneities, probably due
to local variations of the optical properties of the grains. Color maps
strengthen sub-structures eventually present in the envelopes as in the case of V\,Hya. In this
star, the "hottest point" indicated by the color maximum, does not coincide with the center
of the envelope and the direction defined by these two points is practically 
orthogonal to the equatorial disk observed in the CO maps (Sahai et al. 2003).
It is worth mentioning that this direction coincides approximately with 
the jet seen in [SII] emission. The color map of Roberts\,22 indicates
the presence of bipolar lobes, with north-east
side being "hotter" than the south-west side (Fig.\,\ref{color_temp_r22}).

Temperatures near the central regions (angular distances $\theta\,\sim\,1\arcsec$) are
in the range 300\,-\,350\,K whereas at $\theta\,\sim\,4\,\arcsec$ the values are in the
range 250\,-\,280\,K, if we assume grains with typical sizes of about 0.15\,$\mu$m.
Larger grains  ($r_g\,\sim\,5\,\mu$m) give higher values: 430\,-\,525\,K
at $\theta\,\sim\,1\,\arcsec$ and 300\,-\,350\,K at $\theta\,\sim\,4\arcsec$. Comparison
with previous studies, assuming similar grain properties, produce consistent
results. 

The optical properties of amorphous carbon grains measured in laboratory by
different experiments are not exactly the same and were characterized by
Models 1, 2 and 3, according to the authors who have performed the
measurements. The temperature depends on the {\it ratio} between the
grain absorption efficiency $Q_{\lambda}$ at two (nearby) 
wavelengths (see Eq. 4). Thus, the resulting values are not
significantly altered according to the adopted
set of measurements of the grain properties. This is not the case for the grain 
density, which depends directly on $Q_{\lambda}$ (Eq. 3). As a consequence,
densities derived from Model 1 are always  larger than those derived
from Models 2 and 3, if small grains ($r_g\,\sim\,0.1\,\mu$m) are considered. For 
larger grains ($r_g\,\sim\,5\,\mu$m) the differences are not significant. 

Our calculations are based on the hypothesis that the dust envelope
is optically thin at wavelengths around 10\,$\mu$m. The consistency of this
assumption was verified {\it a posteriori}, since optical depths are indeed
less than one and the computed spectrum in the interval 8\,-\,12\,$\mu$m is
consistent with IRAS-LRS data, in the cases  of V~Hya and
IRC\,+10216. The estimated 
mass-loss rate for V\,Hya is compatible with that
derived by Kahane et al. (1996), if grain dimensions of  about 0.2\,$\mu$m
are adopted and one uses the optical Models 1 or 2. Model 3, with the same
grain size, leads to a mass-loss rate a factor of three smaller. However, we
notice that for a grain size of 0.2\,$\mu$m, Knapp et al. (1997) derived a
mass-loss rate one order of magnitude higher than our values. A similar 
analysis for IRC\,+10216 leads to the conclusion that, for distances of about
200\,pc, our mass-loss rates are in agreement, in particular, with the determinations of
Keady et al. (1988) and Groenewegen (1997)  if grain sizes in the range
0.05-0.16~$\mu$m are adopted, independent of optical grain model.

\begin{acknowledgements}
We thank the referee, H. Olofsson, for suggestions that led to
  significant improvements of this paper. We would also like to thank
  H. Bourdin for providing his {\it dispima} graphics
  software. E.L. would also like to thank S. Peirani
  for instructive conversations about ``ISSA NISSA''. 
\end{acknowledgements}


\end{document}